# New group of stable icosahedral quasicrystals: structural properties and formation conditions

Tsutomu Ishimasa*, Yasushi Kaneko and Hiroshi Kaneko

Division of Applied Physics, Graduate School of Engineering,
Hokkaido University, Kita-ku, Sapporo 060-8628, Japan

Abstract

Structural studies on the icosahedral quasicrystals in Zn-Mg-Sc, Cu-Ga-Mg-Sc, and Zn-Mg-Ti alloys as well as their corresponding 1/1 cubic approximants, have revealed that these quasicrystals belong to a new structural group similar to Cd-based quasicrystals. This group is characterized by a triple-shell icosahedral cluster different from both Mackay- and Bergman-types. The presence of the atomic cluster has been deduced from the structure model of the approximant crystal, $Zn_{17}Sc_3$, in which the clusters are embedded in a periodic network of so-called 'glue atoms'. Density measurement suggested the presence of at least 2.7 Zn atoms in the first shell of the cluster in this approximant. The substitutional relationship in these three quasicrystals indicates the important role of Hume-Rothery rule for the formation of this type of quasicrystal. The occurrence of a P-type icosahedral quasicrystal in Zn-Mg-Yb alloy is also reported.

* Corresponding author: Tsutomu Ishimasa
Division of Applied Physics, Graduate School of Engineering, Hokkaido Univ., Kita-ku, Sapporo 060-8628, Japan
Telephone: +81-11-706-6643
Fax: +81-11-706-6643

e-mail address: ishimasa@eng.hokudai.ac.jp

## 1. Introduction

After the discovery of the Al-Mn quasicrystal by Shechtman et al. [1], many icosahedral quasicrystals were found in various alloy systems, including thermodynamically stable phases, such as Al-Cu-Fe, Al-Pd-Mn, Zn-Mg-L (L: lanthanoid elements), etc. In particular, in the past two years, new icosahedral quasicrystals have been discovered in Cd-based [2,3,4], Mg-based [5], Zn-based [6,7], Ag-In based [8] and Cu-based [9] alloys. All quasicrystals known so far are classified into several groups from metallurgical and structural points of view. Classification with respect to the local atomic configurations (or atomic clusters) is useful to understand the local structural features of an icosahedral quasicrystal. Two famous atomic clusters are the so-called Mackay and Bergman clusters [10], in which the atoms are arranged to satisfy the icosahedral symmetry. Usually, Al-based quasicrystals such as Al-Cu-Fe and Al-Pd-Mn are classified as Mackay-type, and Zn-based quasicrystals such as Zn-Mg-L as Bergman-type. The existence of approximant crystals including these two types of clusters [10], in which the clusters are arranged in periodic manner, may support this classification. However, recent studies on the approximants suggested that the classification by only two clusters is too simple [11]. Moreover, the Cd-based quasicrystals were expected to have an atomic cluster which is different from the Mackay and Bergman types [4]; this was deduced from the structures of their approximants.

The stability of alloy phases has frequently been discussed in terms of the Hume-Rothery rule. According to this rule, a certain type of structure is stabilized at a fixed valence electron concentration, $e/a$, by the interaction between the Brillouin zone and the Fermi surface. It is believed that in the case of quasicrystals the Mackay-type forms approximately at $e/a$=1.75 and the Bergman-type at approximately 2.1 [12]. Therefore, it is interesting to study the applicability of the Hume-Rothery rule in the alloys studied here.

The purpose of this article is to review the formation conditions and the structural similarities of the following three icosahedral quasicrystals recently discovered by our group, Zn-Mg-Sc [6], Cu-Ga-Mg-Sc [9] and Zn-Mg-Ti [7]. Some additional information on the approximant crystals will be presented. The role of the Hume-Rothery rule for the formation of these quasicrystals will also be discussed.

## 2. Experimental procedure

The details of the preparation method of Zn-(Mg)-Sc alloys were described in Refs. 6 and 7, and those of Cu-Ga-(Mg)-Sc in Ref. 9. The method to produce the Zn-Mg-Ti alloys is described here. Weighed materials of Zn (Nilaco, 99.998%), Mg (Nilaco, 99.95%) and Ti (Nilaco, 99.9%) were enclosed in a package made of pure molybdenum foil (Nilaco, 99,95%, thickness 0.05 mm) to avoid evaporation and chemical reaction with a silica ampoule. The package was sealed into the silica ampoule in an atmosphere of 200 Torr argon after previous evacuation to a pressure of $2.5 \times 10^{-6}$ Torr. The alloying process of Zn-Mg-Ti tends to be rapid and extreme like an explosion, probably due to its large heat of reaction. If such explosion occurs, it is impossible to control the alloy composition. Such trouble can be avoided by shaking the ampoule several times in the course of the elevation of the temperature. The specimen was kept at 700 °C for 0.25 h for melting, and then quenched in water without breaking the ampoule. The quenched specimens were subsequently annealed at 477-519 °C for 52-108 h. By this procedure, the weight loss during the melting and annealing processes was able to be kept within 0.5 %. However, the present Zn-Mg-Ti specimens tend to lack macroscopic uniformity in the ingots based on systematic changes in the X-ray diffraction patterns. This tendency is pronounced in the alloys that form the icosahedral quasicrystal.

Structural characterization was carried out using transmission and scanning electron microscopy, as well as the powder X-ray diffraction method. The details of the characterization methods are described elsewhere [6]. No contamination was detected from the molybdenum package in both the Zn-Mg-Sc and Zn-Mg-Ti alloys.

The mass density was measured by Archimedes' method using powdered specimens. In particular, the density of the $Zn_{85}Sc_{15}$ alloy was measured very carefully using powdered specimens in the following three different ranges of particle diameter; 150-250, 250-355, and 355-400 μm. The negligible differences of the measured values of the density for each particle size indicated that the effect of small voids existing in the bulk ingot was reduced.

## 3. Formation of new icosahedral quasicrystals

### 3.1. Zn-Mg-Sc icosahedral quasicrystal

The Zn-Mg-Sc quasicrystal forms near the alloy composition of $Zn_{81}Mg_4Sc_{15}$ [6,7]. The powder X-ray diffraction pattern of the $Zn_{80.5}Mg_{4.2}Sc_{15.3}$ alloy, annealed at 750 °C, is presented in Fig. 1(a). Almost all of the diffraction peaks are indexed to the P-type (primitive type) icosahedral quasicrystal, with a 6-dimensional lattice parameter $a_{6D}$=7.111±0.001 Å. Typical electron diffraction patterns of the Zn-Mg-Sc quasicrystals are presented in Figs. 2(a) and (b). The observation of the 5-fold and 2-fold axes are clear evidence of the formation of the icosahedral quasicrystal. The presence of $\tau^3$-scaling in the 2-fold diffraction pattern indicates a P-type quasilattice, where $\tau$ denotes the golden mean. The number of reflections and their sharpness indicate a high quality of the quasicrystal. The positions of the reflections in these patterns satisfy the icosahedral symmetry with high accuracy. The widths of the reflections, $\Delta g_{//}=\Delta q_{//}/2\pi$, (full width at half maximum) measured in the X-ray diffraction pattern range between 1.1 and $1.4\text{x}10^{-3}$ Å$^{-1}$ in the range of $g_{\perp}$ between 0.02 and 0.16 Å$^{-1}$. These widths indicate that the correlation length is larger than at least 700 Å. No clear dependence of the peak widths on the magnitude of $g_{\perp}$ was detected. These observations indicated that the quality of the Zn-Mg-Sc icosahedral quasicrystal is high. The density of the $Zn_{80.5}Mg_{4.2}Sc_{15.3}$ quasicrystal was measured to be 6.17±0.03 g/cm$^3$.

The quasicrystal sometimes exhibits external habit planes perpendicular to the 2-fold axis, and has the shape of a triacontahedron, as presented in Fig. 3(a). This specimen was synthesized in the $Zn_{81}Mg_4Sc_{15}$ alloy by the following heat treatment; the weighed materials were melted at 819 °C, and cooled to 648 °C with a cooling rate of 57 K/h. Recently, it has been shown that the millimeter sized single-quasicrystal of Zn-Mg-Sc forms by the slow-cooling method [13]. Thus, it is expected that this quasicrystal is a congruent or nearly congruent melting phase.

### 3.2. Cu-Ga-Mg-Sc icosahedral quasicrystal

The Cu-Ga-Mg-Sc quasicrystal forms near an alloy composition of $Cu_{48}Ga_{34}Mg_3Sc_{15}$ as a thermodynamically stable phase [9,14]. The X-ray diffraction pattern presented in Fig. 1(b) indicates the presence of a P-type icosahedral quasicrystal with $a_{6D}$=6.938±0.001 Å. The quality of the Cu-Ga-Mg-Sc quasicrystal is as high as that of Zn-Mg-Sc based on the peak widths $\Delta g_{//}$, 1.1-1.5x$10^{-3}$ Å$^{-1}$ measured in the powder X-ray diffraction pattern. A highly symmetric arrangement of reflections can be seen in the electron diffraction patterns presented in Figs. 2(c) and (d). The density of the $Cu_{48.2}Ga_{33.8}Mg_{3.0}Sc_{15.0}$ alloy was measured to be 6.77±0.02 g/cm$^3$.

### 3.3. Zn-Mg-Ti icosahedral quasicrystal

The Zn-Mg-Ti quasicrystal forms in the alloys annealed at relatively low temperature below 484 °C, and has a composition of approximately $Zn_{84}Mg_9Ti_7$. While this quasicrystal has good stability approximately below 484 °C, it has not been clarified yet whether it is thermodynamically stable. The powder X-ray diffraction pattern presented in Fig. 1(c) indicates the presence of the icosahedral quasicrystal with a 6-dimensional lattice parameter, $a_{6D}$=7.033±0.002 Å, in the $Zn_{84.0}Mg_{9.0}Ti_{7.0}$ alloy annealed at 484 °C for 108h. Although the Zn-Mg-Ti quasicrystal exhibits a structural similarity to the other two quasicrystals, this quasicrystal has two special features [7]. The first is the appearance of weak F-type ordering in the 2-fold electron diffraction pattern as can be seen in Fig. 2(f). The degree of ordering changes from place to place, and sometimes only diffuse maxima were observed. The F-type ordering is so weak that it is difficult to identify the order reflections in the powder X-ray diffraction pattern presented in Fig. 1(c). The second feature is a relatively low structural quality compared with the other two quasicrystals. The deviation of the positions of the weak reflections from the exact symmetrical positions is frequently observed in the electron diffraction patterns, which indicates the presence of phason strain [15]. An example of the deviation is indicated by the arrowhead in Fig. 2(f). The peak widths in the X-ray diffraction pattern are relatively wide; for example, the $02\bar{2}\bar{3}03$ reflection has a peak width $\Delta g_{//}$=2.0x$10^{-3}$ Å$^{-1}$ , which is 1.7 times larger than that of the Zn-Mg-Sc quasicrystal. All efforts to search in alloy composition and annealing temperature to improve both the quality and uniformity have not been successful yet. Although this quasicrystal has such internal imperfection, a beautiful external habit with a dodecahedral shape was observed in some specimens of the Zn-Mg-Ti alloy. Two examples of the dodecahedral shape are presented in Fig. 3(b). Although the formation of sub-millimeter sized single quasicrystal is an indication of stability of the quasicrystal, further examination is necessary to clarify this.

## 4. Structure properties of approximant crystals

All three quasicrystals described above have 1/1 cubic approximants near their alloy compositions. Fig. 4 summarizes the powder X-ray diffraction patterns of these approximants. The lattice parameters of the approximants in the $Zn_{85.5}Sc_{14.5}$, $Cu_{49.0}Ga_{35.8}Sc_{15.2}$, and $Zn_{79.0}Mg_{14.6}Ti_{6.4}$ alloys were 13.843±0.001, 13.47±0.01 and 13.531±0.001 Å, respectively. These lattice parameters are similar to each other, and can be related to the 6-dimensional lattice parameters of the corresponding quasicrystals, since all are 1/1 approximants. Electron diffraction experiments revealed that these approximants have the same diffraction symmetry $m\bar{3}$, which is a subgroup of the icosahedral symmetry, $m\bar{3}\bar{5}$. While the Zn-Sc and Cu-Ga-Sc cubic structures have body-centered, I-type, Bravais lattices, the Zn-Mg-Ti has a primitive type. Moreover, the former two approximants have a very similar intensity distribution in the powder X-ray diffraction patterns, as can be seen in Figs. 4(a ) and (b). The Zn-Mg-Ti has a slightly different intensity distribution from other two approximants, and has weak reflections with indices having an odd sum. Electron diffraction experiments indicated that the space group of the Zn-Mg-Ti approximant is P23, $P2_13$, $Pm\bar{3}$, $Pn\bar{3}$ or $Pa\bar{3}$. The difference in Bravais lattice types just corresponds to the difference of quasilattice types in these three icosahedral quasicrystals, which is consistent with the lattice-quasilattice relation reported in the Al-based alloys [16]. Except for this difference, the structural properties of these approximants are similar to each other. The Zn-Sc and Cu-Ga-Sc approximants are assigned, respectively, to the $Zn_{17}Sc_3$ and $Cu_{3.7}Ga_{2.3}Sc$ phases reported in the literatures [17,18]. On the other hand, there is no other report on the 1/1 Zn-Mg-Ti cubic approximant to our knowledge.

The structure model of the $Zn_{17}Sc_3$ crystal is presented in Fig. 5, which was analyzed by Andrusyak et al. [17] by means of single X-ray diffraction method. This structure can be interpreted as a body-centered cubic (bcc) arrangement of atomic clusters with triple shells; the first shell of a dodecahedron consists of Zn, the second shell of an icosahedron of Sc, and the third shell of an icosidodecahedron of Zn. The so-called glue Zn atoms form a cage as presented in Fig. 5(d). The triple-shell cluster fits perfectly in this polyhedron. Therefore the $Zn_{17}Sc_3$-type structure can be regarded as a combination of a periodic cage network of glue atoms and the embedded clusters.

Furthermore, it is possible to relate the cage to a distorted triacontahedron with edge-centered Zn atoms, as presented in Fig. 5(e). A very similar structure is also found in $Cd_6Yb$ [19] and $Cd_6Y$ [20] crystals, which are interpreted as 1/1 approximants of the Cd-based quasicrystals, as Tsai et al. pointed out [4]. While the structure model of the Zn-Sc crystal proposed by Andrusyak et al. [17] has a hole in the first shell, the models of the Cd-based crystals have four Cd atoms.

We have checked the presence of atoms in the first shell by measuring the mass density of the $Zn_{85}Sc_{15}$ alloy, which is almost single phase in the cubic approximant, containing only a small amount of an impurity phase of the $Zn_{58}Sc_{13}$-type. The measured density was 6.47±0.02 g/cm$^3$. This value indicates the presence of 2.7 Zn atoms per shell, estimated from the density and the measured lattice parameter, $a$=13.843Å, assuming that other sites are occupied by 24 Sc and 136 Zn with a 100% occupancy. Therefore, the exact stoichiometry of the Zn-Sc approximant is expected to be not $Zn_{85}Sc_{15}$, but $Zn_{85.5}Sc_{14.5}$. Actually a entirely single-phase sample was obtained at this alloy composition. It is interesting to note that similar calculation for the $Cd_6Yb$ crystal suggests the same value, 2.7 Cd atoms, using the density, 8.70 g/cm$^3$, and $a$=15.638Å given in Ref. 19. These structural similarities between the Zn-Sc and the Cd-based approximants strongly indicate that the three quasicrystals, Zn-Mg-Sc, Cu-Ga-Mg-Sc and Zn-Mg-Ti, belong to the same structure type as the Cd-based quasicrystals.

These approximants are key to solving the structure of this type of icosahedral quasicrystal. The triacontahedron presented in Fig. 5 (e) has an edge length of 5.03 Å in the Zn-Sc approximant, which is exactly equal to the edge length of the Penrose rhombohedra, calculated from the lattice parameter, $a_{6D}$=7.111 Å, of the Zn-Mg-Sc quasicrystal. It is well known that many triacontahedra are included in the ideal 3-dimensional Penrose tiling. If one regards each triacontahedron in the Penrose tiling as a cage for the triple-shell cluster and places the edge-center atoms on the triacontahedron, one can obtain a skeleton model of the Zn-Mg-Sc quasicrystal. Such ideas may be helpful to solve the structure of this type of quasicrystal.

## 5. Hume-Rothery rule as prerequisite of new icosahedral quasicrystals

These experimental results indicate a structural similarity between the three icosahedral quasicrystals and the Cd-based quasicrystals. These quasicrystals have common features with respect to the constituent elements, as summarized in Table 1: (1) the bases of these alloys are sp-metals such as Mg, Cu, Zn, Ag, Cd, and In, and (2) they also include several percent of Ca or transition metals, such as Sc, Ti, Y or lanthanoid metals, as minor elements. The average numbers of valence electrons par atom, $e/a$, are listed in Table 1. For the calculation of $e/a$, we assumed a valence of 1 for Cu and Ag, 2 for Mg, Ca, Zn, Cd and Yb, and 3 for Sc, Y and the lanthanoid metals, except for Yb, and 4 for Ti. The values of $e/a$ range from 2.00 to 2.15. This near equality suggests a Hume-Rothery mechanism for the stabilization of these icosahedral quasicrystals.

The relations between Zn-Mg-Sc and Cu-Ga-Mg-Sc, and also between Zn-Mg-Sc and Zn-Mg-Ti, are typical examples of stabilization by the Hume-Rothery mechanism. In the former alloy, Zn is replaced by Cu and Ga, and in the latter Sc is replaced by Ti, Mg and Zn, under the condition of approximately constant values of $e/a$. Other evidence of a Hume-Rothery mechanism can be found in the equality of the Fermi wave vector, $k_F$, and one half of a wave vector of a strong reflection. In the case of the Zn-Mg-Sc quasicrystal, $k_F$=0.251 Å$^{-1}$ equal to one half of the wave vector for the $02\bar{1}\bar{2}02$ reflection, $g_{02\bar{1}\bar{2}02}/2$=0.2511 Å$^{-1}$. Here, the Fermi wave vector was calculated for an electron concentration $e/a$=2.15, using the measured density, ρ=6.17 g/cm$^3$, by assuming a free electron gas model. A similar equality can be found also in the Zn-Sc approximant crystal. These suggest that the presence of the Fermi edge at the pseudo gap is related to the strong reflection. This electronic structure is responsible for the increase of the Pauli paramagnetic susceptibility, measured at high temperature in the Zn-Mg-Sc quasicrystal and the Zn-Sc approximant [21]. However, in the case of the Cu-Ga-Mg-Sc quasicrystal, good agreement is obtained between $k_F$ and $g_{02\bar{1}\bar{2}02}/2$=0.257Å$^{-1}$ using the measured density ρ=6.77 g/cm$^3$, if one uses $e/a$ =2.15 instead of 2.01. This may indicate that the valence of Cu is 1.3 instead of 1.0. This is supported by the fact that the quasicrystals including divalent metals such as Ca and Yb have $e/a$=2.00, but those including trivalent metals have $e/a$ equal to 2.15, except for the Cu-Ga-Mg-Sc quasicrystal. Therefore, we believe that the valence of Cu is 1.3 in the Cu-Ga-Mg-Sc quasicrystal. However, it is a question why two slightly different values of $e/a$, namely 2.00 and 2.15, correspond to one unique structure in Table 1.

Ishii and Fujiwara have proposed the role of sp-d hybridization in the stabilization of the Cd-based and Zn-Sc approximants [22]. The combination of sp-metals and transition metals found in Table 1 indicates the additional role of sp-d hybridization to the Hume-Rothery mechanism.

Very recently the formation of a P-type icosahedral quasicrystal was noticed in the Zn-Mg-Yb alloy. The electron diffraction patterns of the quasicrystal are presented in Fig. 6. This quasicrystal was formed as a minor phase in the alloy with a nominal composition of $Zn_{75}Mg_{14}Yb_{11}$. Further study is necessary to confirm the formation of the Zn-Mg-Yb quasicrystal.

## 6. Conclusion

The present study has revealed the formation of icosahedral quasicrystals in the following three alloy systems: Zn-Mg-Sc, Cu-Ga-Mg-Sc and Zn-Mg-Ti. The Zn-Mg-Sc and Cu-Ga-Mg-Sc quasicrystals are equilibrium phases. These three quasicrystals have alloy compositions satisfying approximately constant values of $e/a$, similar 6-dimensional lattice parameters, and similar intensity distributions in powder X-ray diffraction patterns. They also have 1/1 cubic approximants with similar structures. The density measurement revealed the presence of at least 2.7 Zn atoms in the first shell of the Zn-Sc approximant. These common features strongly indicate that they belong to a unique structure type, into which Cd-based quasicrystals are also classified. The role of the Hume-Rothery mechanism is important as a prerequisite of the formation of these quasicrystals. However, we still do not have a sufficient condition for their formation. The role of Mg atoms in Zn-Mg-Sc and Cu-Ga-Mg-Sc may give us a useful hint on the sufficient condition, because the formation of the quasicrystals is quite sensitive to the presence of a small amount of Mg.

## Acknowledgment

The authors would like to thank R. Maezawa and T. Mitani for their contributions to the specimen preparation and X-ray diffraction experiments. This work was supported by a Grant-in-Aid for Scientific Research B from Japan Society for the Promotion of Science.

Figure Captions

Fig. 1 Powder X-ray diffraction patterns of (a) $Zn_{80.5}Mg_{4.2}Sc_{15.3}$, (b) $Cu_{48.2}Ga_{33.8}Mg_{3.0}$ $Sc_{15.0}$, (c) $Zn_{84.0}Mg_{9.0}Ti_{7.0}$ alloys measured by Cu K$\alpha$ radiation. Open and full triangles in the lower margin indicate the reflections due to the 1/1 cubic approximant and $Zn_{22}Ti_{3}$-type phase, respectively.

Fig. 2 Electron diffraction patterns of the Zn-Mg-Sc icosahedral quasicrystal in (a) and (b), the Cu-Ga-Mg-Sc one in (c) and (d), and the Zn-Mg-Ti quasicrystal in (e) and (f). An arrow in (b) indicates the $11\bar{1}\bar{1}12$ reflection with a lattice spacing d=2.374 Å. An example of the F-type superlattice reflections is indicated by an arrow in (f). Notice the distortion of the small rhombus, indicated by an arrowhead in (f).

Fig. 3 Scanning electron micrographs of (a) the triacontahedron of Zn-Mg-Sc quasicrystal and (b) the dodecahedron of Zn-Mg-Ti quasicrystal.

Fig. 4 Powder X-ray diffraction patterns of (a) $Zn_{85.5}Sc_{14.5}$, (b) $Cu_{49.0}Ga_{35.8}Sc_{15.2}$, (c) $Zn_{79.0}Mg_{14.6}Ti_{6.4}$ alloys measured by Cu K$\alpha$ radiation. Miller indices of bcc and primitive-type cubic structures are inserted in (a) and (c), respectively.

Fig. 5 Structure model of the 1/1 Zn-Sc cubic approximant after Andrusyak et al: (a) dodecahedron of Zn; (b) icosahedron of Sc; (c) icosidodecahedron of Zn; (d) arrangement of glue Zn atoms; (e) triacontahedron of Zn with edge-center Zn atoms.

Fig. 6 Electron diffraction patterns of the P-type icosahedral quasicrystal formed in a Zn-Mg-Yb alloy: (a) 5-fold ; (b) 2-fold diffraction patterns.

Table 1

Icosahedral quasicrystals classified into a new group. In $Cd_{65}Mg_{20}L_{15}$, L denotes the rare-earth metals Y, Gd, Tb, Dy, Ho, Er, Tm and Lu.

| Alloy | *e/a* | Reference |
|---|---|---|
| $Cd_{84}Yb_{16}$ | 2.00 | [3,4] |
| $Cd_{85}Ca_{15}$ | 2.00 | [3,4] |
| $Cd_{85-x}Mg_xYb_{15}$ $0 \leq x \leq 60$ | 2.00 | [5] |
| $Cd_{85-x}Mg_xCa_{15}$ $0 \leq x \leq 50$ | 2.00 | [5] |
| $Ag_{42}In_{42}Yb_{16}$ | 2.00 | [8] |
| $Ag_{42}In_{42}Ca_{16}$ | 2.00 | [8] |
| $Cu_{48}Ga_{34}Mg_3Sc_{15}$ | 2.01 | [9] |
| $Cd_{65}Mg_{20}L_{15}$ | 2.15 | [2] |
| $Zn_{81}Mg_4Sc_{15}$ | 2.15 | [6] |
| $Zn_{84}Mg_9Ti_7$ | 2.14 | [7] |

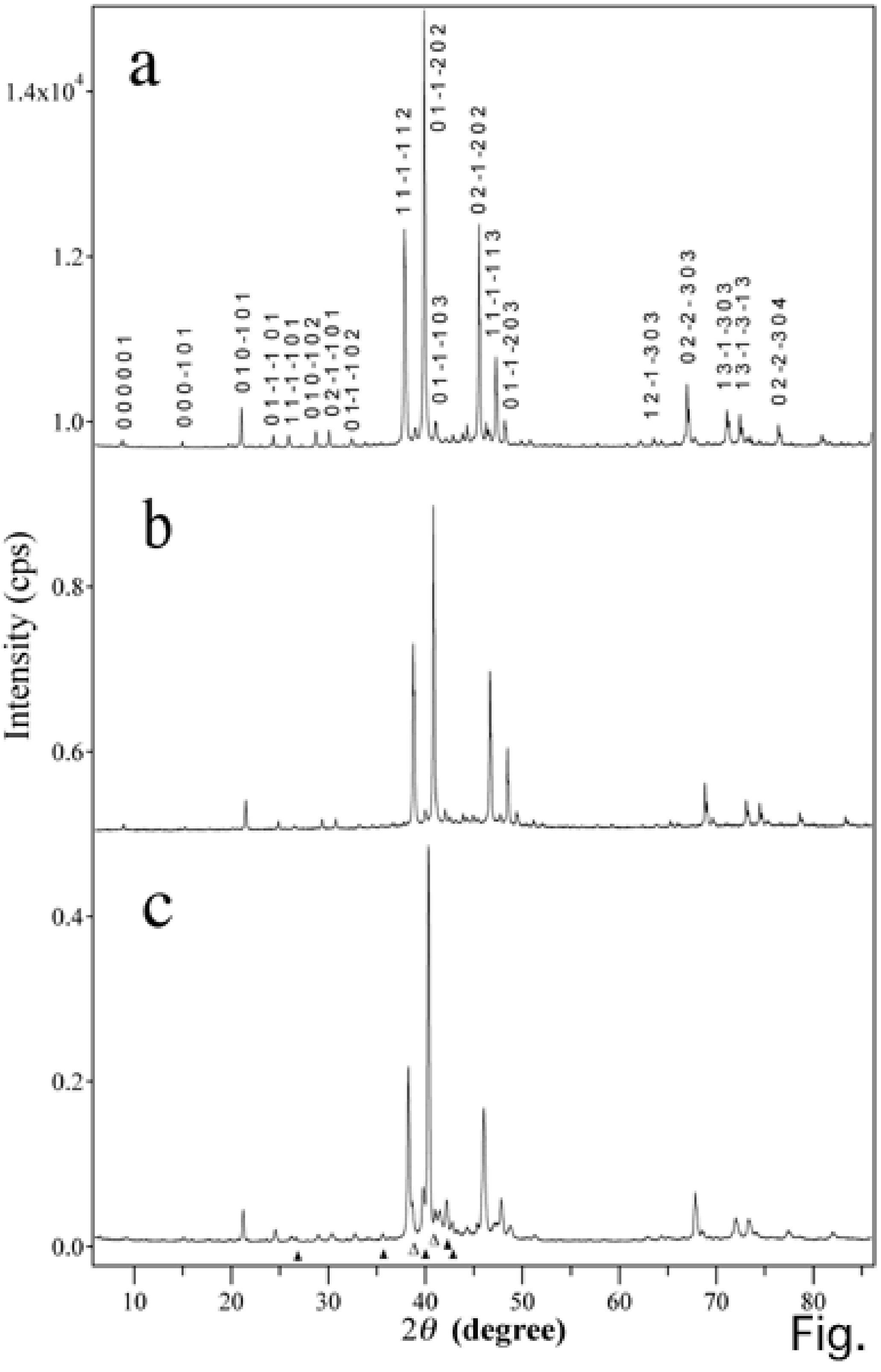

Fig. 1

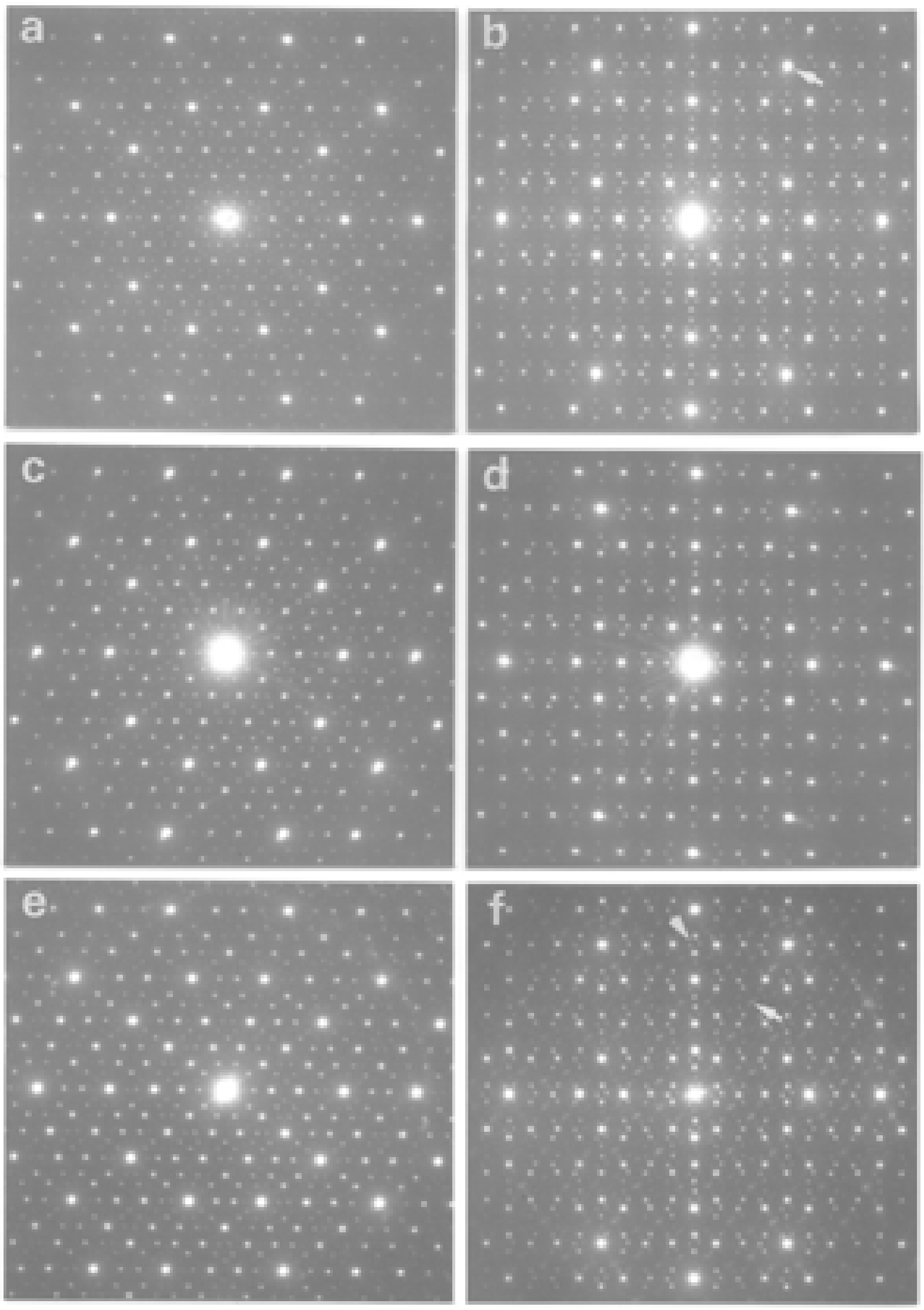

Fig. 2

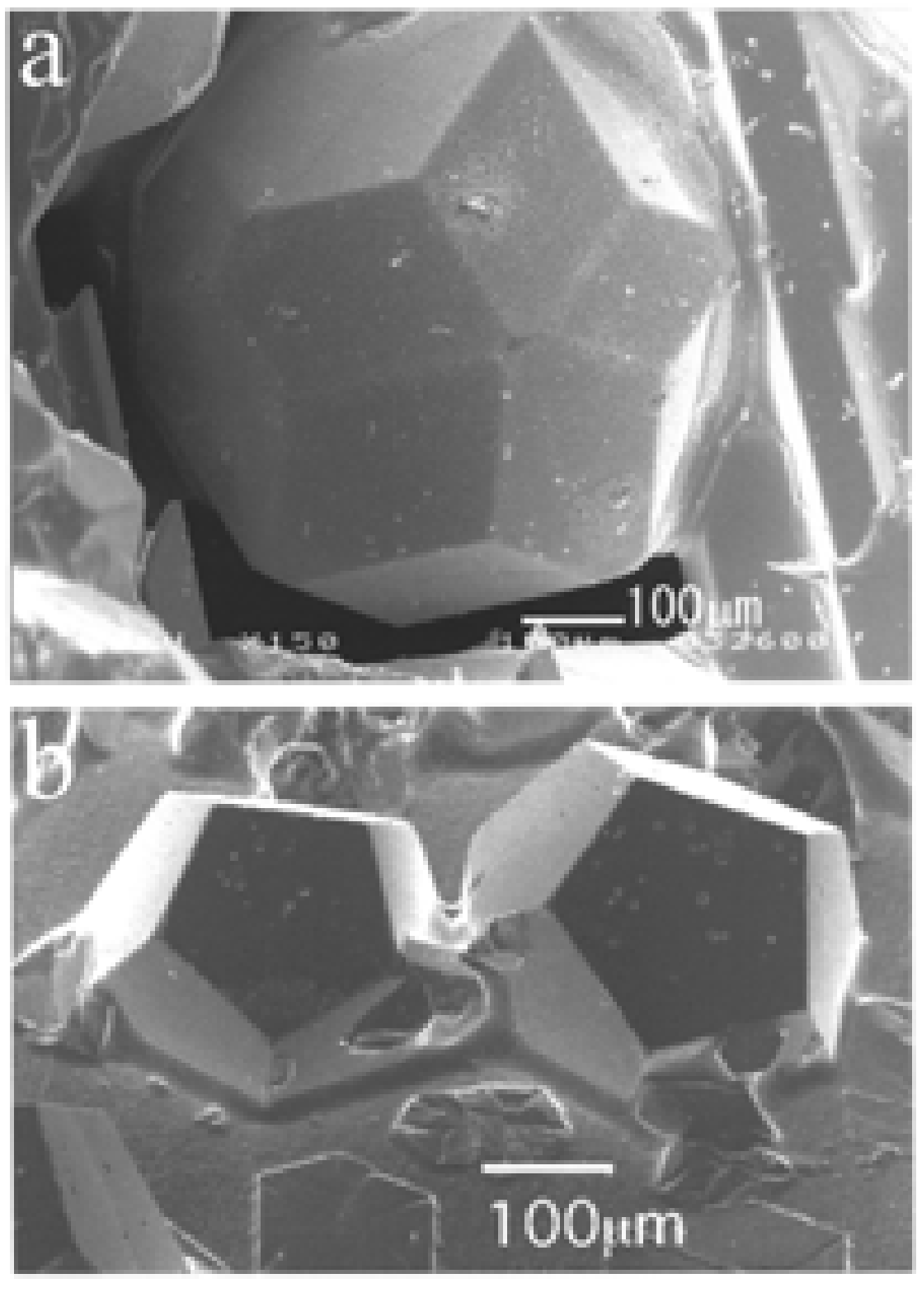

Fig. 3

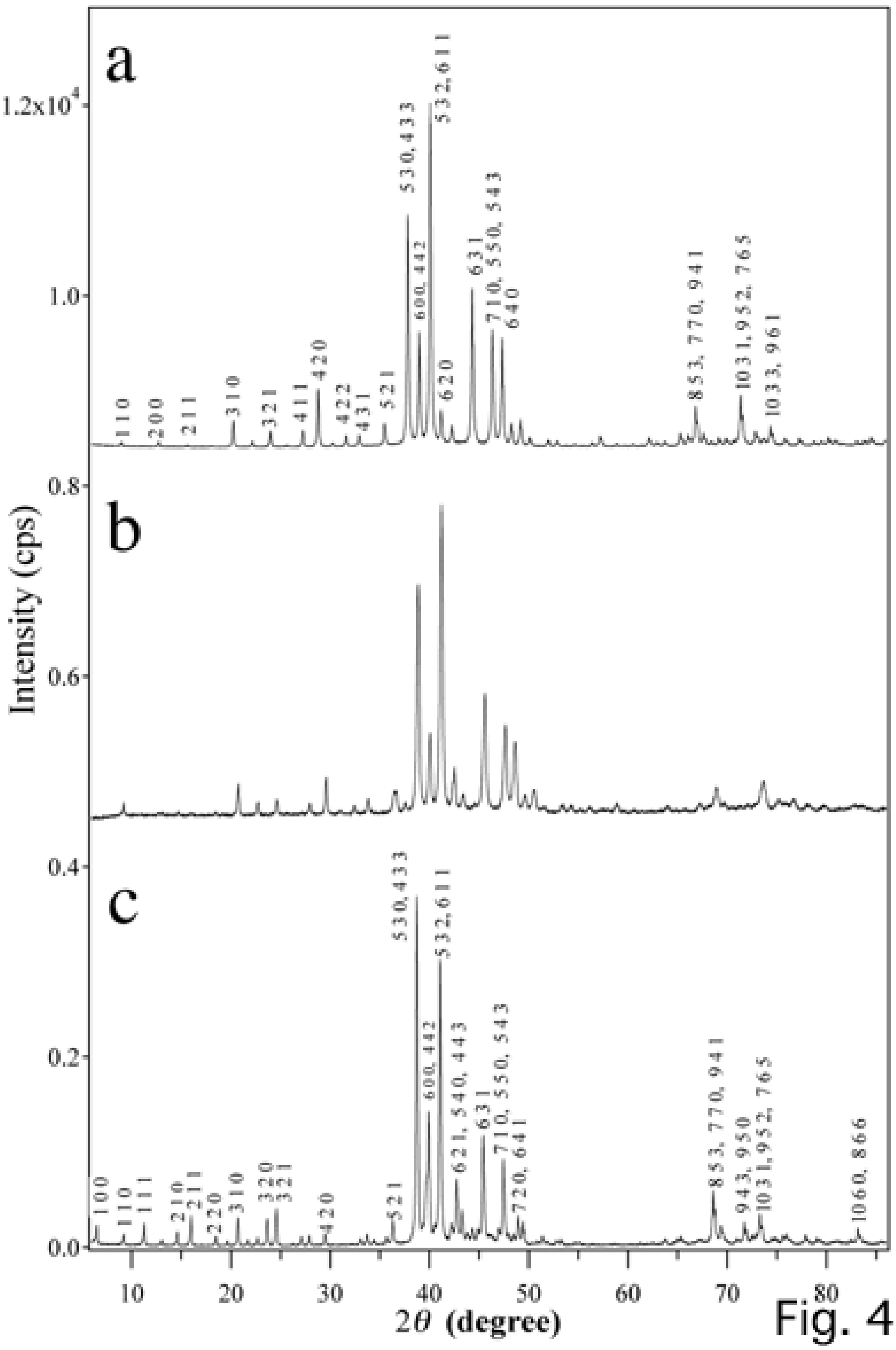

Fig. 4

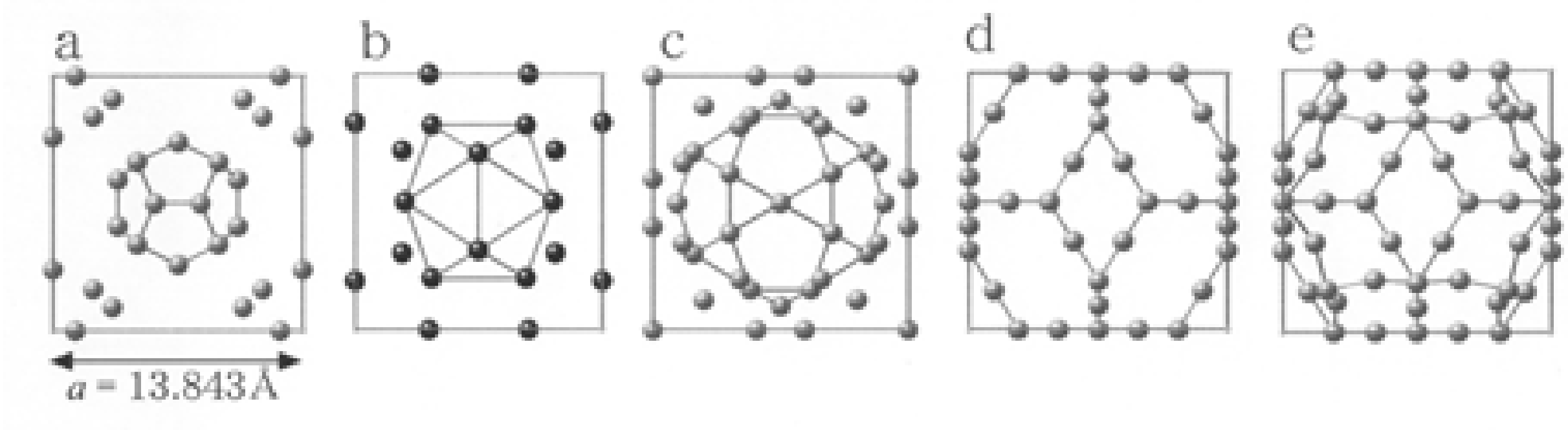

Fig. 5

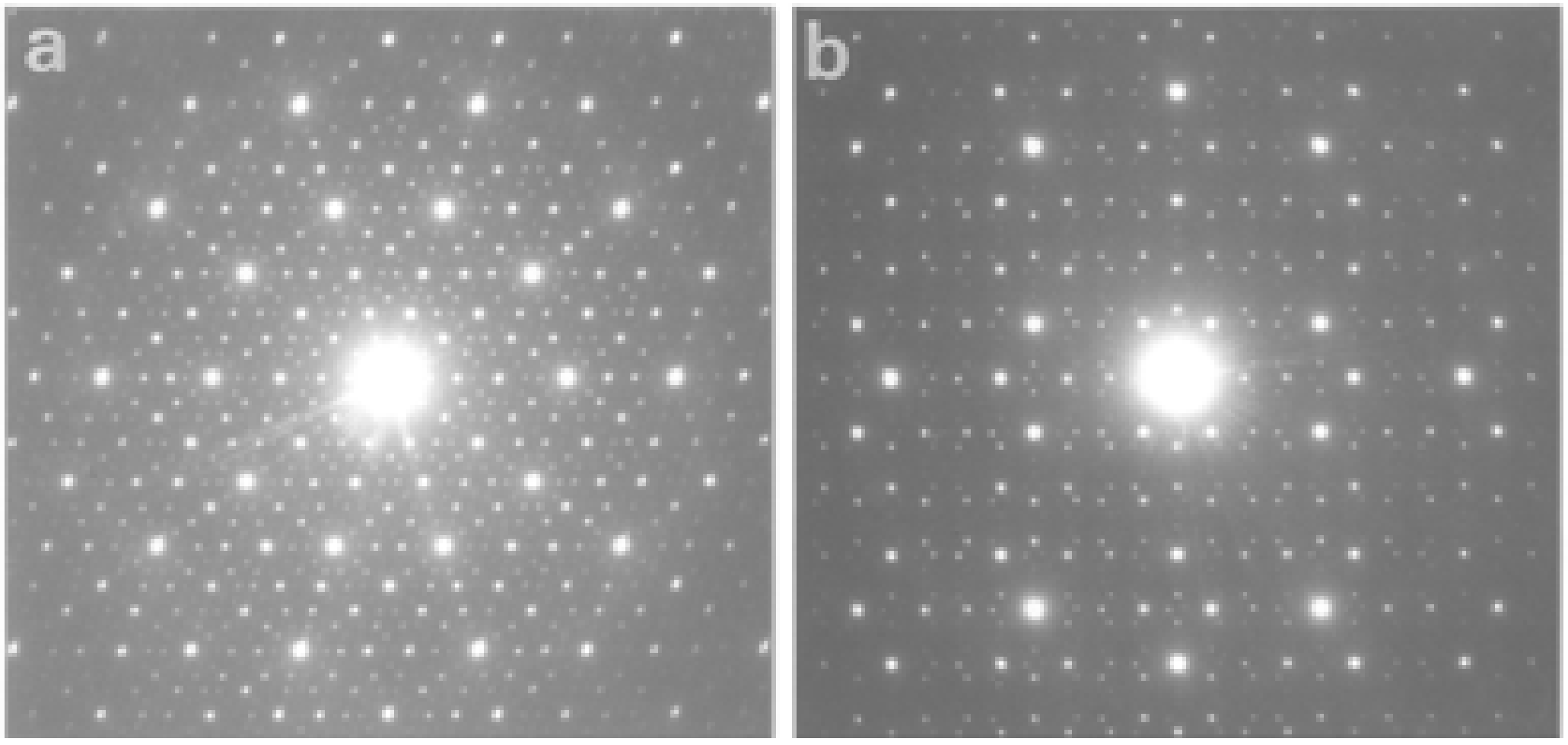

Fig. 6